# Al$_2$O$_3$-CaO macroporous ceramics containing hydrocalumite-like phases

O.H., Borges [(a)*]; T., Santos Jr [(a)]; V. R., Salvini [(b)]; V.C., Pandolfelli [(a)]

[(a)] Federal University of São Carlos, Graduate Program in Materials Science and Engineering.
[(b)] São Paulo State Technological College (FATEC), Rua Jordão Borghetti 480, Sertãozinho, SP, 14160-050, Brazil

*Corresponding author at: +55-16-33518253; fax: +55-16-33615404.
E-mail: otavio.borges@dema.ufscar.br

**Abstract**

A mechanism to explain the lower onset strengthening temperature induced by CaCO$_3$ in alumina-based macroporous ceramics is proposed, which relies on hydrocalumite-like phase formation during processing. Close to 600 °C, such phases are decomposed to lime and mayenite (12CaO·7Al$_2$O$_3$), where the latter, due to its intrinsic nanoporosity and high thermal reactivity, generates bonds between the ceramic particles at ~700ºC, resulting in microstructure strengthening. Based on this premise, the authors concluded that other Ca$^{2+}$ sources could act similarly. Indeed, compositions containing Ca(OH)$_2$ or CaO showed the same effect on the onset strengthening temperature, which reinforces the proposed mechanism. The results attained indicated that macroporous insulators could be thermally treated at lower temperatures, just to acquire enough mechanical strength for installation, finishing *in-situ* their firing process. Besides that, lower sintering temperatures could be used to produce macroporous ceramics that would be applied in low thermal demand environments, e.g. aluminum industries.

# 1  Introduction

The development of new technologies has always been closely linked to energy consumption. This trend and the fact that the world main power supplies are non-renewable and highly polluting, has resulted in increasing concern about the energy issue [1–3]. According to the International Energy Agency (IEA), the industrial sector plays a key role in energy demand, accounting for 37 % of the total amount spent in 2017 [4]. In addition, special attention should be given to industrial high temperature processes (above 1300 ºC), as they are responsible for one-third of the industrial power demand [5]. In this case, much of the energy consumed is used to heat, adjust or maintain the temperature of different equipment, such as furnaces and boilers. Thus, thermal insulators, which reduce heat losses, can help these processes by lowering the energy costs and helping the environment [1,4,5].

Various refractory thermal insulators were developed for this purpose. Among them, those of greater commercial importance are currently comprised by ceramic fibers [6,7]. Even though these materials are easily available and present low thermal conductivity at room temperature, they have a short lifecycle due to the high surface area of fibers and the presence of $SiO_2$-containing binders, which might undergo densification in service. Moreover, these insulators can be highly toxic when inhaled and some compositions are classified as carcinogenic by the International Agency for Research on Cancer (IARC) [8,9].

Therefore, the search for novel non-toxic thermal insulators able to withstand high temperatures with a longer lifecycle is of great scientific and technological interest. One of the options that emerges as a promising alternative is the macroporous refractories [6]. By definition, they comprise a refractory matrix containing pores with a diameter greater than 50 nm, which are present in a volumetric fraction above 40% [10,11]. This microstructure is able to halt both the heat transference by phonons, due to the discontinuities in the solid crystalline lattice (pores), and by radiation, as pores in the range of 0.5 μm to 3 μm can interact with electromagnetic waves in the infrared spectra, spreading them [12].

Thus, some macroporous refractories exhibit similar values of effective thermal conductivity when compared to fibrous blankets and bricks, having the added advantage of being non-toxic and presenting a superior lifecycle, as they do not densify in service, when previously fired [6,13]. However, there are some challenges to overcome if macroporous refractories are to be more widely used to replace fiber-based insulators. For example, this material needs to be pre-sintered at high temperatures in order to achieve suitable mechanical strength, even if its application is for lower thermal demand environments.

Recently, it was reported that by using $CaCO_3$ in alumina-based dense refractories, besides inducing hibonite formation ($CaO.6Al_2O_3$, also referred to as $CA_6$) that helps to reduce materials' thermal conductivity, and it is also capable of decreasing the strengthening onset temperature [14]. Investigating this effect, Luz *et al.* [14] reported the elastic modulus increase of dense alumina-based refractory castables as a function of temperature, by changing the $CaCO_3$ - CAC ratio (see Figure 1). However, the mechanism by which this effect takes place has not yet been explained. It is worth highlighting that Lin and Shen [15] described an effect called Sintering-Coarsening-Coalescence (SCC) on $CaCO_3$ particles exposed to temperatures close to 400 °C, but the reasons why refractories containing $CaCO_3$ and $Al_2O_3$ showed E increase at lower temperatures remain unclarified.

Additionally, the Layered Double Hydroxide (LDH) phase formation in refractories processed in the presence of water has been studied with promising results, so far [16]. LDHs with hydrocalumite structure can be generated in systems containing $Al^{3+}$ and $Ca^{2+}$ ions in aqueous media [17], which is comprised by alternated stacks of calcium and aluminum hydroxide layers, with their interlamellar spaces filled in with charged species [17]. Hydrocalumite-like phases present the following chemical formula $[Ca_2Al(OH)_6]^+ X^- \cdot mH_2O$, where X is the anion located in the interlamellar space counterbalancing the positive charges, *e.g.* $OH^-$, $CO_3^{2-}$ and $Cl^-$ [18,19]. This hydrate phase decomposes close to 600 °C, resulting in lime (CaO) and mayenite ($12CaO.7Al_2O_3$ or $C_{12}A_7$) [17].

Moreover, it is known that mayenite has intrinsic nanometric porosity resulted by the nanosized cages present in its microstructure, which are partially filled by weakly bounded $O^{2-}$ [20,21]. This feature classifies mayenite as an anti-zeolite phase. Recent studies pointed out that mayenite, besides releasing peroxide ions from 350 °C to 650 °C, reducing the number of filled cages, undergoes phase transition close to 650 °C, giving rise to a phase in which the lattice parameter increases more rapidly with the temperature [20,22].

These data might induce the correlation between the effect reported by Luz *et al.* [14] for dense refractories with those related to hydrocalumite-like phase formation at the interface among the ceramic particles, which results in CaO and $C_{12}A_7$ during thermal treatment. In addition, this hypothesis can predict that other $Ca^{2+}$ sources will also be able to generate such an effect, anticipating the onset strengthening temperature, as long as they form hydrocalumite-like phases during the processing.

In order to analyze this hypothesis, the hydrocalumite-like phase formation during the macroporous alumina-based insulator processing, containing both $CaCO_3$ and $Ca(OH)_2$, was evaluated. Following that, the elastic modulus *versus* temperature profile ($E_{\text{ in situ}}$) for compositions containing distinct $Ca^{2+}$ sources was carried out.

## 2  Materials and methods

### 2.1  Materials

To investigate the LDH formation for compositions containing $CaCO_3$ and $Ca(OH)_2$, samples with hydrocalumite stoichiometric quantities of $Al^{3+}$ and $Ca^{2+}$ were produced in order to maximize their formation, and thus enable their detection. One composition presented $CaCO_3$ as $Ca^{2+}$ precursor (RHI Magnesita, Brazil, $D_{90}$ = 12.64 μm, $D_{50}$ = 7.38 μm, $D_{10}$= 1.49 μm ), whereas the other contained $Ca(OH)_2$ (Synth, Brazil, 99.9% purity, $D_{90}$ = 11.00 μm, $D_{50}$ = 5.66 μm, $D_{10}$= 2.76 μm) . In both, alumina CL 370 (Almatis, Germany, $D_{50}$ = 2.5 μm) was used as an $Al^{3+}$ source. Ceramic powders were added to a beaker with distilled water to result in a suspension with 20 wt% of solids and kept under stirring for 30 minutes at room temperature. Finally, the samples were dried firstly at 50 ºC for 24 h and then at 110 ºC for another 24 h, generating a white powder.

Besides these samples, specially prepared to identify the hydrocalumite-like phase formation in the systems tested, formulations of macroporous ceramics were produced to investigate the effect of distinct $Ca^{2+}$ sources on the earlier Young modulus increase and other properties. These compositions comprised a liquid foam containing alumina and different binders and $Ca^{2+}$ sources. The foam formulation and the additives evaluated in this study are presented in Table 1 and Table 2, respectively. It should be emphasized that the equivalent percentage of CaO was kept constant (1.7 wt%) in order to generate 20 wt% of $CA_6$ in the samples under thermodynamic equilibrium.

Macroporous ceramics were prepared by direct foaming technique [23], by incorporating a previously prepared liquid foam into an alumina-based suspension. Alumina CL370 and CT3000SG (Almatis, Germany) were used in the appropriate proportion to obtain maximum packaging according to the Andreasen packing model (q = 0.37). Alumina suspensions were prepared with distilled water, adding Castament® FS60 (BASF, Germany) and a non-ionic surfactant based on alkylpolyethylene glycol ether (Lutensol® AT 50 Pulver, BASF, Germany) as dispersing agents.

The liquid foam was prepared with a commercial liquid surfactant based on 1,2-benzisothiazol-3(2H)-one (Vinapor GYP 2680, BASF, Germany) and the thickening additive Hydroxyethyl Cellulose (Cellulosize® 100 CG FF, Dow, USA). Hydratable alumina (Alphabond 300, Almatis, Germany) or calcium aluminate cement (CAC, Secar 71, Imerys Aluminates, France) were used as binders. Besides CAC, $CaCO_3$, $Ca(OH)_2$ and CaO (obtained from $CaCO_3$ calcination at 900 °C for 4 hours and presenting $D_{90}$ = 7.97 μm, $D_{50}$ = 3.24 μm, $D_{10}$ = 1.36 μm) were used as $Ca^{2+}$ sources.

Table 1: Alumina ceramic foam composition.

|  | Components | wt% |
|---|---|---|
| **$Al_2O_3$ Suspension** | CL370 | 76.06 |
|  | CT3000SG | 5.72 |
|  | Castament® FS60 | 0.09 |
|  | Lutensol AT50 | 0.1 |
|  | Water | 15.79 |
| **Foam** | Vinapor GYP 2680 | 3.4680 |
|  | Hydroxietil Cellulose | 0.0056 |

Table 2: Binders and lime sources used for each ceramic foam formulation.

| | | Additive (wt%) | | | | |
|---|---|---|---|---|---|---|
| | | Secar 71 | AlphaBond 300 | $CaCO_3$ | $Ca(OH)_2$ | CaO |
| **Formulation** | REF | - | 1.00 | - | - | - |
| | 5 % CAC | 5.00 | - | - | - | - |
| | 0.53 % $CaCO_3$+CAC | 4.00 | - | 0.53 | - | - |
| | 2.04 % $CaCO_3$+CAC | 1.00 | - | 2.04 | - | - |
| | 1.55 % $Ca(OH)_2$+CAC | 1.00 | - | - | 1.55 | - |
| | 1.42 % Lime | - | - | - | - | 1.42 |

For processing the samples, initially FS60 was dissolved in water, followed by adding the aluminas and Lutensol AT50. Then, the system was maintained under intense mechanical stirring for ~ 30 minutes. After that, the additives (Table 2) were incorporated to each composition, except for CAC and calcined CaO ones, which, due to their faster hydration kinetics, had to be added at the last processing step. A liquid foam was produced by mechanical stirring Vinapor GYP 2680 and Cellulosize® 100 CG FF. After preparing the foam, it was mixed with the alumina suspension under minimal stirring (30 rpm). Finally, the liquid foam containing ceramic particles was cast into cylinder (50 mm x 50 mm) and bar (150 mm x 25 mm x 25 mm) molds. Compositions REF and 1.42 % Lime were cured at 50 °C for 24 h and dried at 110 °C for another 24 h. The other compositions containing CAC were cured in a climatic chamber at 50 ºC and humidity of 80 % for 24 hours, then dried at 110 °C for 24 h.

*2.2 Methods*

The samples specially prepared to enable hydrocalumite-like phase detection were analyzed by Fourier transform infrared diffraction spectroscopy (DRIFT) using Equinox 55 equipment (Brüker, Karlsruhe, Germany). The samples were scanned with wavelengths in the range between 400 nm and 4000 nm with a resolution of 0.4 nm and 32 scans per wavelength. In addition, DRIFT scans were carried out for the raw materials used to produce the evaluated suspensions, in order to compare their transmittance spectra.

X-ray diffraction (XRD) was performed to investigate hydrocalumite-like phase formation. Due to difficulties of identifying these phases by XRD of green samples, they were calcined at 700 ºC for 2 hours, allowing the decomposition of the likely present hydrocalumite. Therefore, the detection of mayenite in these samples (which in systems containing CaO and $Al_2O_3$ is expected to be formed close to 1300 ºC [24]) would indicate the presence of hydrocalumite-like phases in the initial systems. XRD analyses were carried out in Geiger-Flex diffractometer (Rigaku, Tokyo, Japan) in the 4 º to 90 º range, with 0.02 º step and using copper Kα radiation and nickel filter. The results were compared to diffraction patterns from the International Diffraction Data Center (ICDD) database.

The macroporous ceramic foams were characterized by their elastic modulus *versus* temperature profiles ($E_{in\ situ}$) in Scanelastic 02 (ATCP Physics Engineering, Brazil), using the resonance frequency scanning method according to ASTM E1875 [25]. Elastic moduli were measured every 5 minutes up to 1400 ºC and back to room temperature, with a heating and cooling rate of 2 ºC.min$^{-1}$. Additionally, cold crushing strength (ASTM C133-97 [26]) and linear shrinkage were measured for the 2.04 % $CaCO_3$+CAC composition in their green condition (after drying at 110 °C for 24 h) and after firing at 400 °C, 600 °C, 800 °C and 1000 °C for 5 hours.

Mineralogical analysis of macroporous ceramic samples were carried out to evaluate whether the predicted $CA_6$ content would actually be achieved after firing at

1600 ºC for 5 hours. Thus, fired samples were ground and the powder was analyzed by X- ray diffraction at the previously described conditions. The obtained patterns underwent Rietveld refinement using Topas software (Bruker, Massachusetts, USA) and ensuring Goodness of Fit (GOF) < 1.5 and weighted profile R-factor (RWP) < 15 %.

## 3  Results and Discussion

*3.1  Diffuse reflectance infrared Fourier transform spectroscopy (DRIFT)*

Samples containing alumina and $CaCO_3$ or $Ca(OH)_2$, named CL370+$CaCO_3$ and CL370+$Ca(OH)_2$, respectively, were characterized by DRIFT. Figure 2 shows the spectra for each sample, their precursors and that of hydrocalumite, as reported by Tian and Guo in [17]. It can be observed that CL370+$CaCO_3$ and CL370+$Ca(OH)_2$ presented almost all characteristic bands of hydrocalumite. In this regard, absorption band at 3640 $cm^{-1}$ represents the stretching vibrations of lattice water together with surface-absorbed $H_2O$. The radiation absorption close to 3485 $cm^{-1}$ is related to $OH^-$ stretching vibrations in the portlandite-like layers of hydrocalumite, whereas in the range of 1640 $cm^{-1}$ it is assigned to the H-O-H bending vibrations of the interlayer water. Infrared waves close to 1445 $cm^{-1}$ induce vibrations for the adsorbed $CO_2$ bonds.

Figure 2

Moreover, the absorption bands at 789 $cm^{-1}$, 533 $cm^{-1}$ and 428 $cm^{-1}$ for the hydrocalumite are related to stretching vibrations of $OH^-$ groups linked to metal elements [17,27]. Therefore, despite not showing some absorption bands in the range of 789 $cm^{-1}$ and 428 $cm^{-1}$, the DRIFT results pointed out that phases with hydrocalumite structure were formed in both CL370+$CaCO_3$ and CL370+$Ca(OH)_2$ compositions. However, the other bands presented in these samples, but not in hydrocalumite, highlighted that not all reactants were consumed to form the LDH phase. Additionally, it is expected that adding calcined CaO will also result in hydrocalumite formation, as it reacts with water generating $Ca(OH)_2$ [28].

The presence of hydrocalumite-like phases in the characterized samples was also confirmed by the XRD, as discussed as follows.

*3.2    X-ray diffraction (XRD)*

Figure 3 shows X-ray profiles for samples of $CL370+CaCO_3$ and $CL370+Ca(OH)_2$ after thermal treatment at 700 °C for 2 h. The diffraction peaks show the presence of corundum, lime and mayenite in both compositions highlighted above.

Mayenite in the $CaO-Al_2O_3$ system is expected above 1300 ºC, therefore its presence at 700 ºC reinforces the hypothesis of hydrocalumite formation, as this phase decomposes close to 600 ºC resulting in CaO and $C_{12}A_7$ [17]. Additionally, it can be observed that not all initial materials reacted to form hydrocalumite-like phases. These results are in agreement with those obtained by DRIFT and suggest that hydrocalumite-like phases could be formed during the processing of macroporous alumina-based refractories containing $CaCO_3$, $Ca(OH)_2$ or CaO as $Ca^{2+}$ sources and produced by direct foaming.

Figure 3

*3.3    In situ elastic modulus of compositions with different $CaCO_3/CAC$ ratios*

Elastic modulus (E) evaluation with temperature was carried out for green samples of REF, 5 % CAC, 0.53 % $CaCO_3$+CAC and 2.04 % $CaCO_3$+CAC compositions in order to analyze the effect of $CaCO_3$ on the initial E increase temperature, as previously reported by Luz *et al*. [14] for dense refractory castables. The results can be seen in Figure 4.

All CAC-containing compositions showed an E drop in the temperature range between 100 ºC and 450 ºC, which is related to the decomposition of hydrated phases [29]. For the REF sample, two E drop temperatures could be seen, one close to 100 ºC and the other about 200 ºC. The first occurs due to the evaporation of free water, whereas the second is related to pseudoboehmite decomposition, an usual phase for hydratable alumina-based refractories [30].

Figure 4

All compositions in Figure 4 present similar initial E values. Nevertheless, after thermal treatment, it can be seen that the REF one showed the highest one at the end of the cycle, which can be explained by its high linear shrinkage (55 % above the average of the other compositions), which resulted in lower porosity. In fact, after firing REF at 1600 °C for 5 h, it showed 25 % less porosity than the $Ca^{2+}$ containing compositions. For $CaCO_3$-containing samples, the higher carbonate content resulted in superior E at the end of the measurements. This behavior could be assigned to morphological differences in the *in situ* formed $CA_6$. When interacting with small alumina particles, $CaCO_3$ favors the generation of acicular hibonite [24], which provides additional strengthening.

Compositions 5 % CAC and REF presented similar temperatures of the initial E increase (950 ºC), which is associated to the beginning of reactive alumina sintering. Thus, it can be concluded that neither CAC nor HA resulted in a significant change in lowering the strengthening temperature. On the other hand, for $CaCO_3$-containing compositions, a correlation was observed between the $CaCO_3$ amount and the initial temperature of E increase (780 ºC and 700 ºC for 0.53 % $CaCO_3$+CAC and 2.04 % $CaCO_3$+CAC, respectively). These results agree with the ones reported by Luz *et al*. [14]. It is worth observing that for $CaCO_3$-based compositions, the expected decrease in the E values during carbonate decomposition (carried out close to 800 ºC) was not observed. This phenomenon, which was also observed for dense refractories [14], could be attributed to the high E increase at this temperature, which overlaps the decrease expected during the $CaCO_3$ decomposition.

Additionally, the correlation between the lower onset strengthening temperature and the presence of hydrocalumite-like phases stands out because, unlike HA and CAC, $CaCO_3$ is able to form these LDHs during the refractory processing [31–34].

### 3.4 Crushing strength and linear shrinkage versus temperature

Moving on to understanding how hydrocalumite-like phases affect the strengthening onset temperature, cold crushing strength and linear shrinkage after thermal treatment were carried out for the 2.04 % $CaCO_3$+CAC samples. This composition was selected because of its higher likelihood of hydrocalumite generation. The samples were fired at 400 ºC, 600 ºC, 800 ºC and 1000 ºC for 5 hours with a heating rate of 2 ºC.min$^{-1}$ and the results are shown in Figure 5.

Figure 5

Compared to the green condition, thermal treatment at 400 ºC and 600 ºC for 5 h did not present significant changes either in crushing strength or in linear shrinkage. On the other hand, samples fired at 800 ºC and 1000 ºC showed a significant cold crushing strength increase (close to 640 % and 1080 %, respectively) without relevant linear shrinkage. These low values of shrinkage indicate that the strengthening mechanism carried out in the range of 600 ºC and 1000 ºC did not involve the classic densification process.

Thus, this work proposes that the observed strengthening mechanism takes place due to hydrocalumite-like phases formed on the particles' surface during the processing stage. As previously mentioned, this LDH phase decomposes around 600 ºC resulting in lime and mayenite [17]. The mayenite formed contains intrinsic nanoporosity and undergoes a phase transition at 650 ºC, which makes it even more reactive [20,22]. Therefore, it is expected that highly reactive mayenite located in some particles` interfaces could react by linking them. As a result, an increase of mechanical strength without shrinkage can be observed before sintering takes place, whereby the latter is expected to occur above 950 ºC.

Looking carefully at Figure 4, it is worth observing that there was no high increase in the E value around 950 ºC for $CaCO_3$-containing compositions, but a constant and smooth increment was noticed. Although further investigations are required, this behavior could be assigned to the particles` links induced by highly reactive mayenite, which lead to mechanical strength values close to those driven by primary chemical bonding after sintering.

### 3.5 *In situ elastic modulus of different $Ca^{2+}$ sources*

After evaluating the mechanical strengthening induced by $CaCO_3$ at lower temperatures, the proposed mechanism and the results from DRIFT and DRX (presented in Figure 2 and Figure 3, respectively), were all used to forecast that an analogous effect should take place by using Ca(OH)$_2$ and CaO sources, instead of CaCO$_3$, as in either cases hydrocalumite-like phases are expected to be formed. To evaluate this prediction, compositions containing calcium hydroxide [1.55 % Ca(OH)$_2$+CAC] and calcium oxide (1.42 % Lime) had their *in situ* E values evaluated. The results are presented in Figure 6, which also reports data for compositions 5 % CAC and 2.04 % CaCO$_3$+CAC.

Figure 6

Initially, the E decrease in the temperature range between 100 ºC and 400 ºC was observed for the 1.55 % Ca(OH)$_2$+CAC composition. This behavior, similar to that presented by 2.04 % CaCO$_3$+CAC, is related to the hydrated CAC phase decomposition and dehydration of Ca(OH)$_2$, in which the latter is expected at temperatures close to 400 ºC [35]. On the other hand, 1.42 % Lime composition showed just one E drop, which can be explained by the absence of CAC and by the CaO hydration during the foam processing, which leads to Ca(OH)$_2$ formation, as previously mentioned. Therefore, the E drop at ~ 400 ºC is caused by Ca(OH)$_2$ decomposition. It is also worth observing that the 5 % CAC composition presented the lowest mechanical strengthening after firing (E$_{final}$ − E$_{initial}$), which is most likely due to the formation of highly asymmetric platelet CA$_6$ grains induced by using CaCO$_3$ and Ca(OH)$_2$ [24].

Analyzing the impact of each Ca$^{2+}$ source on the strengthening, it is clear that CaCO$_3$, Ca(OH)$_2$ and CaO reduced the temperature for initial E increase to 700 ºC, whereas for CAC, this temperature was close to 950 ºC. This behavior reinforces the connection between the hydrocalumite-like phase formation and the lower strengthening onset temperature, as $CaCO_3$, $Ca(OH)_2$ and $CaO$ can develop such phases during the refractory foam processing.

The main advantages on this earlier E increase induced by $Ca^{2+}$ sources can be summarized as: (i) macroporous insulators could be thermally treated at lower temperatures, just to acquire enough mechanical strength for transportation and installation, finishing *in situ* their firing process; (ii) lower sintering temperatures can be used to produce macroporous ceramics that would be applied in low thermal demand environments, e.g. aluminum industry. In both situations, the energy input required to manufacture the macroporous material would be reduced.

Another environmental advantage which stands out is the possibility of replacing, at least in part, the amount of CAC by $CaCO_3$. Calcium aluminate cement manufacturing typically uses $Al(OH)_3$ and $CaCO_3$ as raw materials, which are processed at a high temperature (1550 ºC) [36], whereas $CaCO_3$ is found in nature as lime deposits. Therefore, this replacement would lead to reductions on macroporous refractory carbon footprints.

*3.6    Quantitative phase analysis by XRD*

As previously pointed out, $CA_6$ formation could bring benefits to macroporous insulators, *e.g.* lower thermal conductivity and firing shrinkage. With this in mind, fired samples (1600 ºC for 5 hours) of each formulation had their mineralogical composition quantified by Rietveld Analysis of XRD patterns. The raw patterns and quantitative phase content are presented in Figure 7 and Table 3, respectively.

The expected $CA_6$ amount (20 wt%) was formed after the thermal treatment for all compositions showing that the $Ca^{2+}$ sources tested [$CaCO_3$, $Ca(OH)_2$, $CaO$ and $CAC$] were able to react with alumina at 1600 ºC for 5 hours, resulting in hibonite. Thus, the applied conditions were enough to achieve the thermodynamic equilibrium, leading to the formation of a refractory phase, which can result in higher chemical and dimensional stability to the macroporous ceramic.

Figure 7

Table 3: Quantitative phase analysis of fired ceramic foams with distinct $Ca^{2+}$ sources based on the Rietveld refinement of X-ray diffraction patterns.

| Formulations | Corindum (%wt) | Hibonite (%wt) |
|---|---|---|
| **REF** | 100 | 0 |
| **5 % CAC** | 80.34 | 19.66 |
| **0.53 % $CaCO_3$+CAC** | 78.52 | 21.48 |
| **2.04 % $CaCO_3$+CAC** | 79.61 | 20.39 |
| **1.55 % $Ca(OH)_2$+CAC** | 80.29 | 19.71 |
| **1.42 % Lime** | 80.44 | 19.56 |

However, despite the use of $CaCO_3$, $Ca(OH)_2$ and CaO providing similar benefits to the elastic modulus and mineralogical composition after firing, each $Ca^{2+}$ source can result in different physical properties for the final material. $CA_6$ morphology, mechanical strength, thermal conductivity, linear shrinkage and porosity are some of them. A further paper discussing the impact of distinct $Ca^{2+}$ sources on the physical properties of macroporous refractory ceramics is being prepared and will be submitted soon.

## 4   Conclusions

In this work, the correlation between hydrocalumite-like phase formation during the macroporous ceramic processing and a lower strengthening onset temperature, was studied. Initially, the generation of such phases was observed in systems containing either $CaCO_3$ or $Ca(OH)_2$. Carbonate-containing compositions had their *in situ* elastic modulus evaluated, showing lower strengthening onset temperature. Aiming to understand this phenomenon, crushing strength and linear shrinkage for the 2.04 % $CaCO_3$+CAC composition, fired at distinct temperatures, were measured. From the results obtained, it was observed that the strengthening process indeed started in the 600 ºC – 800 ºC temperature range, without undergoing densification.

Based on the previous results, a mechanism was proposed: Hydrocalumite-like phases generated on the particles' surface during the processing stage, decompose at 600 ºC forming lime and highly reactive mayenite ($12CaO·7Al_2O_3$). This latter phase, located on some particles' interfaces, could react close to 700 ºC linking them and resulting in strengthening without shrinkage. Similar results were also observed for compositions containing either $Ca(OH)_2$ or $CaO$, which support the proposed mechanism, as the selected $Ca^{2+}$ sources are able to form hydrocalumite-like phases during the refractory processing.

Therefore, the use of $CaCO_3$, $Ca(OH)_2$ or $CaO$ could result in macroporous refractory ceramics thermally treated at lower temperatures, which can reduce the energy required to manufacture this type of material. Additionally, replacing part of the CAC content by $CaCO_3$ would lead to lower carbon footprints. Finally, all $Ca^{2+}$ sources tested were able to form the expected $CA_6$ content after firing at 1600 ºC for 5 h and each one of them resulted in different macroporous insulator physical properties, which will be further discussed in a forthcoming paper.

## 5 Acknowledgments

This study was financed in part by the Coordenação de Aperfeiçoamento de Pessoal de Nível Superior - Brasil (CAPES) - Finance Code 001, the Conselho Nacional de Desenvolvimento Científico e Tecnológico – Brasil (CNPq) – Process 130843/2018-0, and Fundação de Amparo à Pesquisa do Estado de São Paulo – Brasil (FAPESP) – Process 2018/07745-5. Additionally, the authors are thankful to BASF Construction Solutions GmbH, to RHI-Magnesita, to Dr. Roger Gonçalves for the DRIFT's analyses and to FIRE – International Federation for Refractory Research and Education.

21**Figure Captions**

Figure 1: Elastic modulus evolution in alumina based castables as a function of temperature and the $Ca^{2+}$ source [14].

Figure 2: Infrared spectra of (a) CL370, $CaCO_3$ and (b) CL370, $Ca(OH)_2$ and the samples produced by combining these raw materials following the steps described in Section 2.1. Both figures show the infrared spectra of hydrocalumite as reported in the literature [17].

Figure 3: X-ray diffraction patterns of compositions containing alumina and $CaCO_3$ or $Ca(OH)_2$ after thermal treatment at 700 °C for 5 hours.

Figure 4: *In situ* elastic modulus as a function of temperature for alumina-based foams with different $CaCO_3$ content.

Figure 5: Crushing strength and linear shrinkage of 2.04 % $CaCO_3$+CAC samples fired at different temperatures. Green samples presented cold crushing strength of (0.51 ± 0.04) MPa and total porosity of (83 ± 2) %.

Figure 6: *In situ* elastic modulus as a function of temperature for alumina-based ceramic foams with different $Ca^{2+}$ sources.

Figure 7: X-ray diffraction patterns of alumina-based ceramic foams fired at 1600 °C for 5 hours. Number 1 represents corindum and number 2 is for hibonite ($CA_6$).

**Figures**

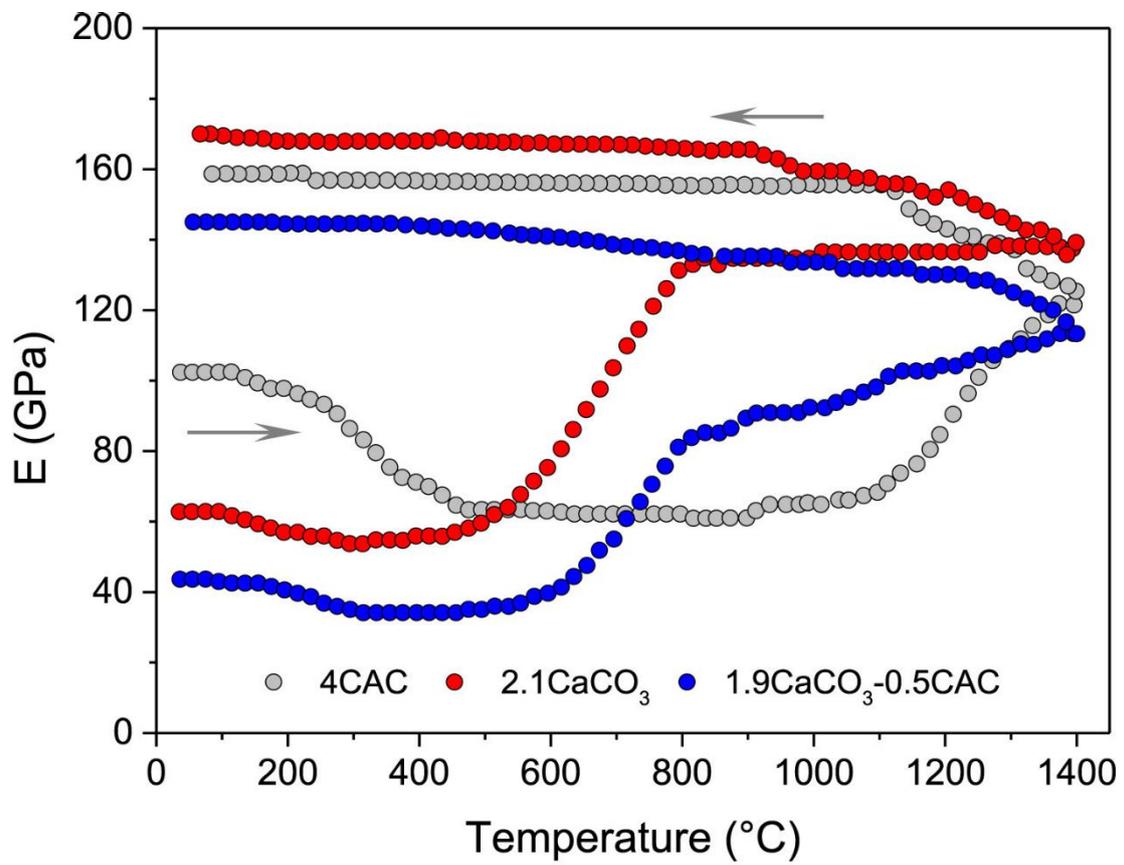

Figure 1: Elastic modulus evolution in alumina based castables as a function of temperature and the $Ca^{2+}$ source [14].

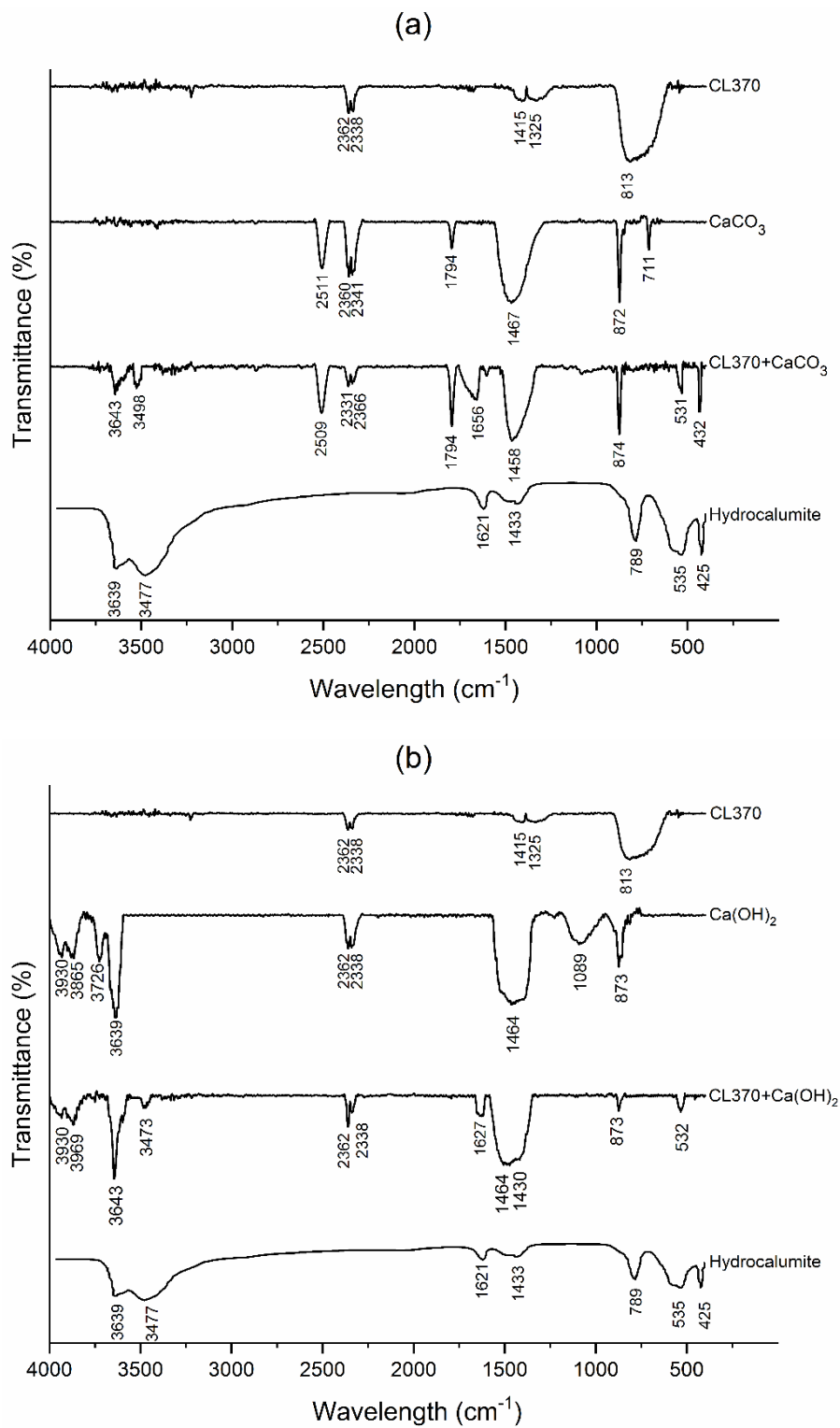

Figure 2: Infrared spectra of (a) CL370, $CaCO_3$ and (b) CL370, $Ca(OH)_2$ and the samples produced by combining these raw materials following the steps described in Section 2.1. Both figures show the infrared spectra of hydrocalumite as reported in the literature [17].

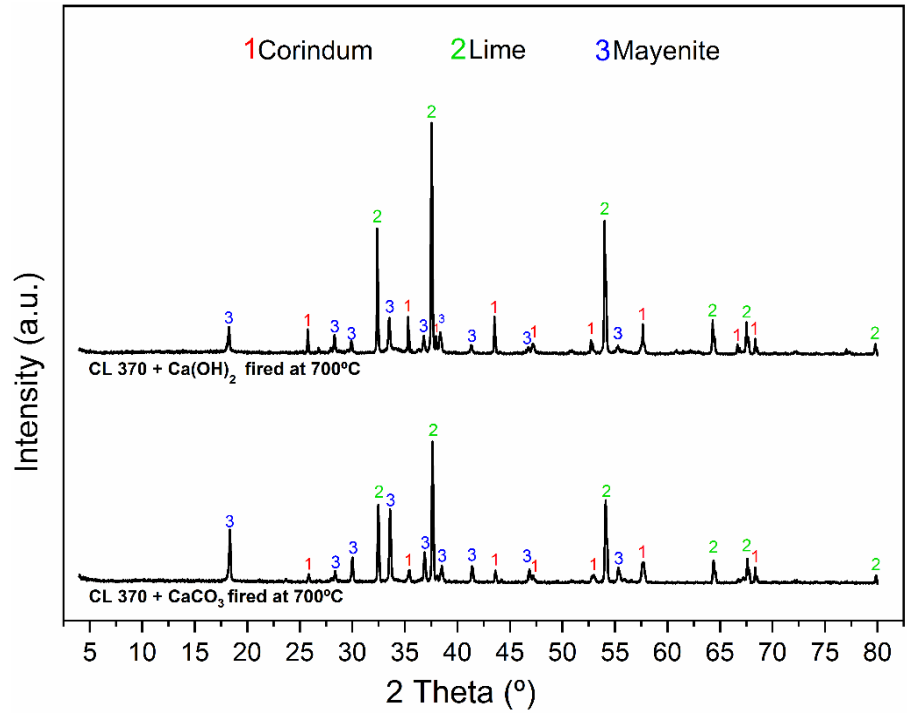

Figure 3: X-ray diffraction patterns of compositions containing alumina and CaCO$_3$ or Ca(OH)$_2$ after thermal treatment at 700 °C for 5 hours.

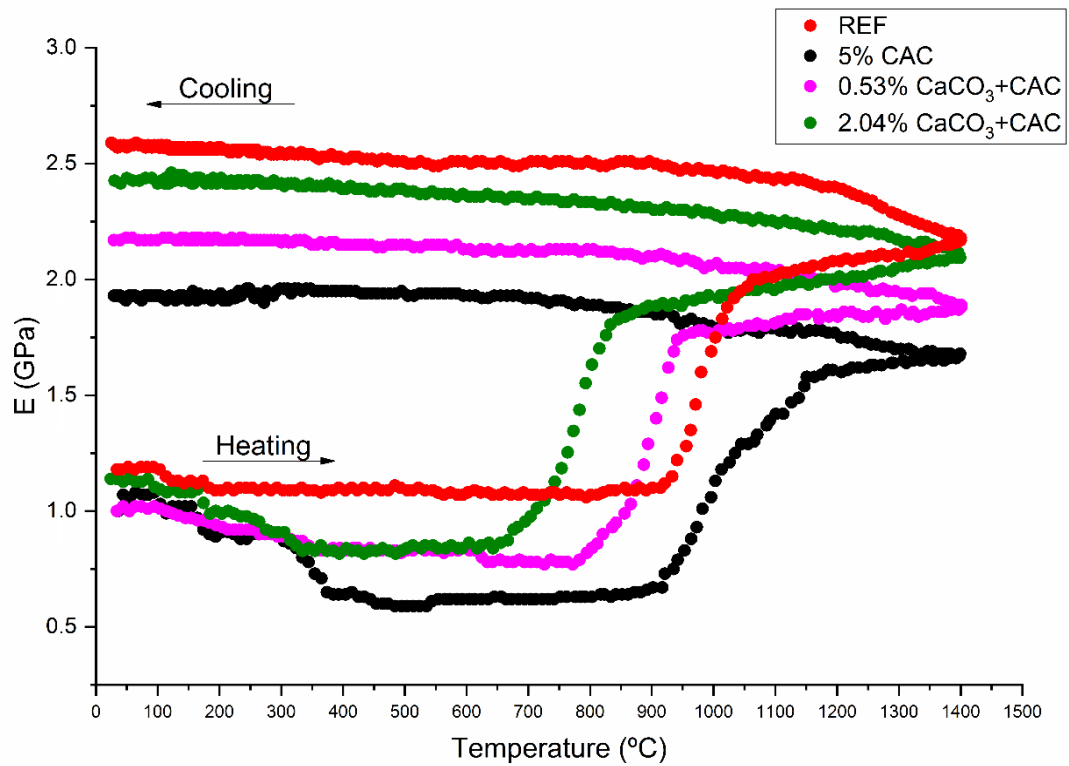

Figure 4: *In situ* elastic modulus as a function of temperature for alumina-based foams with different CaCO$_3$ content.

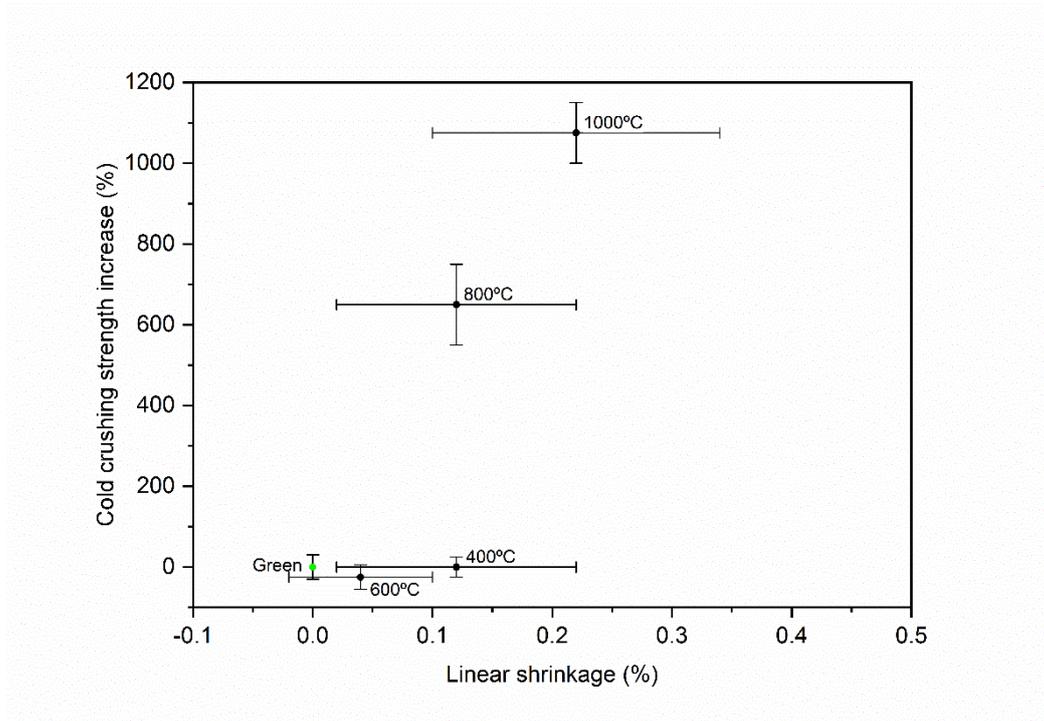

Figure 5: Crushing strength and linear shrinkage of 2.04 % $CaCO_3$+CAC samples fired at different temperatures. Green samples presented cold crushing strength of $(0.51 \pm 0.04)$ MPa and total porosity of $(83 \pm 2)$ %.

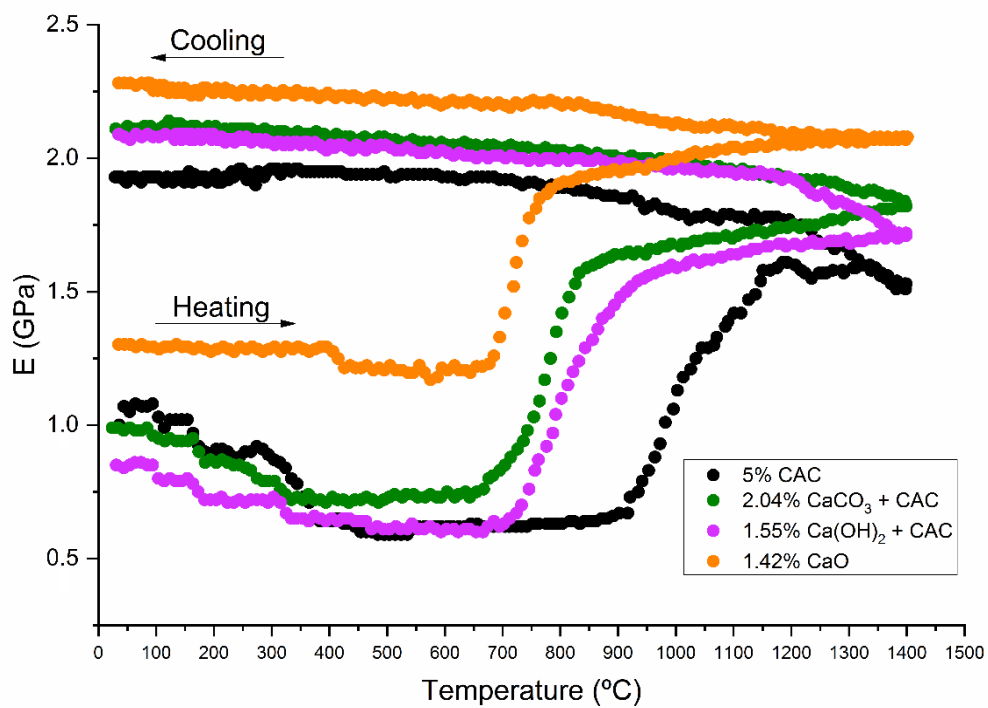

Figure 6: *In situ* elastic modulus as a function of temperature for alumina-based ceramic foams with different $Ca^{2+}$ sources.

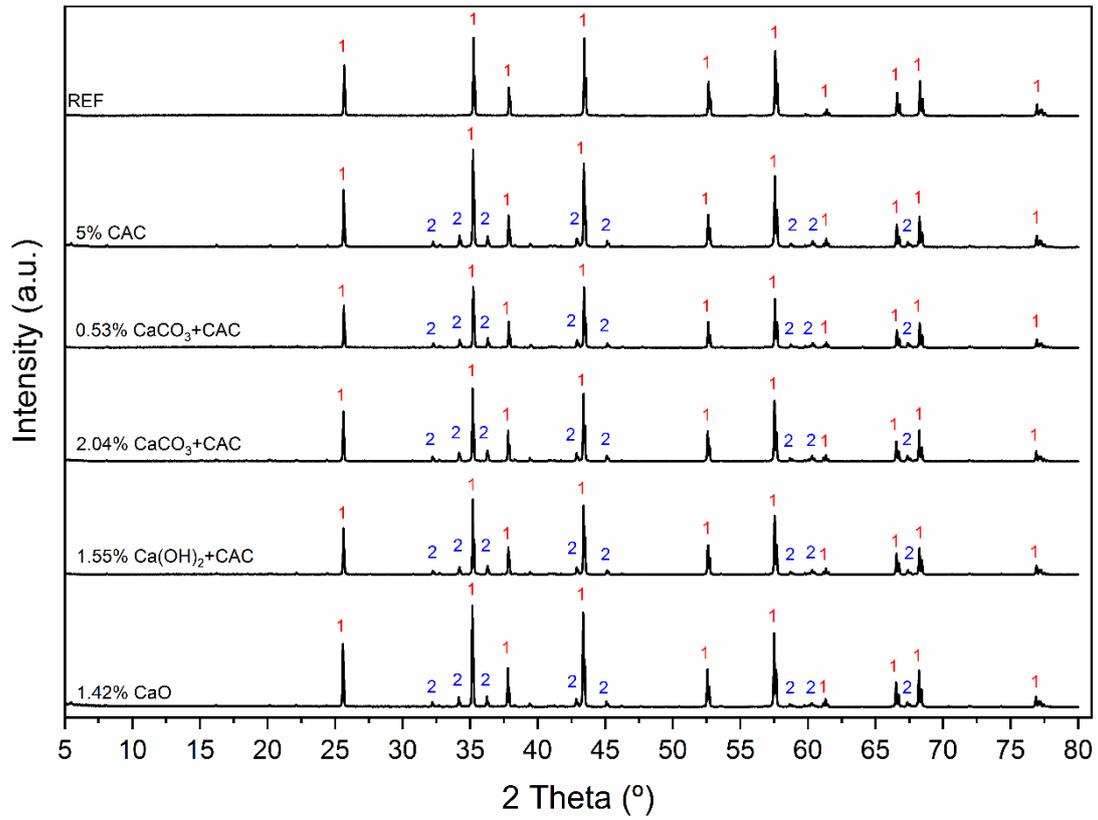

Figure 7: X-ray diffraction patterns of alumina-based ceramic foams fired at 1600 °C for 5 hours. Number 1 represents corindum and number 2 is for hibonite ($CA_6$).